\documentclass[final,english,twocolumn,amssymb,nobibnotes,aps,prb,longbibliography]{revtex4-2}
\usepackage{graphicx}

\usepackage{natbib}
\usepackage{amsmath}
\usepackage{amssymb}
\usepackage{appendix}
\usepackage{comment}
\usepackage[dvipsnames]{xcolor}
\usepackage[section]{placeins}
\usepackage{layouts}
\definecolor{darkblue}{rgb}{0,0,.65}
\definecolor{darkgreen}{rgb}{0.28,0.41,0.19}

\usepackage[pdfauthor={David J. Luitz},pdfstartview=FitH,breaklinks=true,bookmarks=true,colorlinks=true,anchorcolor=black,citecolor=blue,
  filecolor=black,menucolor=black,urlcolor=darkblue,linkcolor=blue,]{hyperref}
\usepackage[all]{hypcap} 
\usepackage{braket}

\newcommand{\tr}{\mathrm{Tr}}

\renewcommand{\vec}[1]{\boldsymbol{#1}}

\newcommand{\E}{\mathrm{e}}

\newcommand{\Tr}{\mathrm{Tr}}
\renewcommand{\Im}[1]{\mathrm{Im}(#1)}
\renewcommand{\Re}[1]{\mathrm{Re}(#1)}

\graphicspath{{figures/}{images/}}

\begin{document}

\title{Many-body Hierarchy of Dissipative Timescales in a Quantum Computer}

\author{Oscar Emil Sommer}
\author{Francesco Piazza}
\author{David J. Luitz}\email{dluitz@pks.mpg.de}
\affiliation{Max Planck Institute for the Physics of Complex Systems, Noethnitzer Str. 38, Dresden, Germany}

\date{\today}

\begin{abstract}
	We show that current noisy quantum computers are ideal platforms for the simulation of quantum many-body dynamics in generic open systems.
	We demonstrate this using the IBM Quantum Computer as an experimental platform for 
	confirming the theoretical prediction from [\href{https://journals.aps.org/prl/abstract/10.1103/PhysRevLett.124.100604}{\textit{Phys. Rev. Lett.} \textbf{124}, 100604 (2020)}] of an emergent hierarchy of relaxation timescales of many-body observables involving different numbers of qubits.	
	
	Using different protocols, we leverage the intrinsic dissipation of the machine responsible for gate errors, to 
	implement a quantum simulation of generic (i.e. structureless) local dissipative interactions.
\end{abstract}

\maketitle

\section{Introduction} 

Quantum many-body systems generically show highly complex
correlations induced by the interactions among their
constituents. This complexity is a major obstacle to their
simulability on classical computers. Quantum simulation is a
promising way to circumvent this problem \cite{feynman1982simulating,georgescu_2014}. A lot of progress has
been made in the last decades using analog quantum simulators. 
While those are specifically built for a given model, their digital
counterparts hold instead the promise for a general purpose quantum
simulation \cite{lloyd1996universal}. There are several examples of successfully implementing quantum
many-body dynamics in digital quantum simulators \cite{o2016scalable,kandala2017hardware,hempel2018quantum,lanyon2011universal,salathe2015digital,barreiro2011open,barends2015digital,langford2017experimentally,martinez2016real}, however, the intrinsic
dissipation present in all platforms severely limits the accessible time scales.

Here, we demonstrate that this disadvantage can be turned into a virtue, and
show that the intrinsic dissipation responsible for gate errors in current quantum computing
platforms can be used as a building block to simulate the physics of
generic open quantum many-body systems.

The term ``generic'' here indicates the absence of any particular
structure, thus covering the vast majority of systems. 
For isolated quantum many-body systems, this means that
states can be classified only by their energy. In this case, generic behavior is well
described within the framework of the eigenstate thermalization hypothesis \cite{deutsch_quantum_1991,srednicki_chaos_1994,srednicki_thermal_1996,oganesyan_localization_2007,rigol_thermalization_2008,borgonovi_quantum_2016,dalessio_quantum_2016,luitz_long_2016,luitz_anomalous_2016,deutsch_eigenstate_2018,luitz_ergodic_2017}, which is
formalized via a proper random-matrix description. On the other hand,
the characterization of the generic behavior of open quantum many-body
systems is still in its beginnings.

Recent work was devoted to setting the foundations of dissipative
quantum chaos \cite{shirai_thermalization_2018,sa_spectral_2019,can_random_2019,can_spectral_2019,luitz_exceptional_2020,denisov_universal_2019,sa2020complex}. By considering completely random models of markovian dissipation, a
characteristic spectrum of the Liouville operator $\mathcal L$ has
been identified \cite{denisov_universal_2019}, which
describes the dynamics of the density matrix $\rho$ of the system in
terms of a quantum master equation $\partial_t \rho = \mathcal
L[\rho]$. These results strongly constrain the range of dissipative
timescales one can expect in the absence of more detailed knowledge on
the system. The assumption of full randomness is however too strong for experimentally relevant situations, since it includes unphysical, nonlocal dissipative interactions.

In a recent theoretical work \cite{wang_hierarchy_2020}, we predicted that, if the dissipation is constrained to be local, a hierarchy of relaxation timescales generically emerges, which reflects the degree of locality of observables. The order of timescales can be predicted quantitatively from a random matrix theory and is nontrivial in the presence of dissipative interactions.

In the present paper, we employ a digital quantum simulator to
experimentally observe this nontrivial hierarchy of dissipative timescales and test the theoretical predictions. Specifically, using the IBM quantum computing platform based on
superconducting qubits \cite{koch_charge-insensitive_2007}, we
exploit the intrinsic dissipation manifest in gate errors to 
implement local dissipative interactions in a controlled manner. 
We then measure the dynamics of a large number of multi-qubit observables to characterize
the hierarchy of decay timescales and to test our theoretical predictions. 

Our approach differs strongly from other currently developed strategies for the simulation of
open quantum systems using appropriately extended sets of unitary gates \cite{wang2011quantum,wang2013,wei2016,candia2015,sweke2015,zixuan2020quantum}. While conventional approaches
assume that gates are essentially perfect and the precision of the simulation suffers from intrinsic gate errors, our approach embraces the imperfections of current platforms and uses them as central building block for the simulation of generic local dissipation.
We demonstrate here that for capturing the generic physics of open
quantum systems, our approach is much cheaper and robust, allowing to observe
non-trivial phenomena of dissipative quantum many-body chaos in currently 
available quantum computing platforms.
Such a deeper understanding of what one can expect in dissipative
quantum many-body systems is also crucial to guide more detailed modeling of decoherence and loss processes, and is in particular important in the context of quantum computing.

\section{Results} \begin{figure}[t]
	\includegraphics{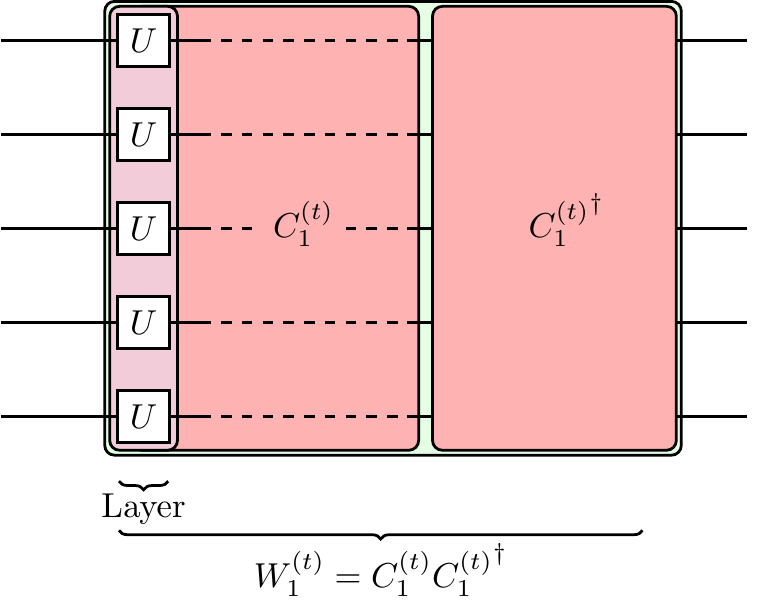}
	\caption{Example of a waiting circuit $W_1^{(t)}$ used to simulate one body
		dissipation. It consists of two subcircuits $C_1^{(t)}$ and the inverse
		circuit ${C^{(t)}_1}^\dagger$. $C^{(t)}_1$ is composed of $t$ layers, each containing $\ell$ different random one body unitary gates $U_3(\theta,\phi,\lambda)$ acting on each qubit. The one body unitaries are sampled from the Haar measure. \label{fig:1body_circuit} }
\end{figure}

We simulate a generic, purely dissipative $\ell$ qubit system, described by a Lindblad master equation, on the IBM Quantum Computing platform.

In Sec. \ref{sec:methods}, we describe in detail the circuits we use to implement generic one body dissipation 
and generic dissipative two body interactions. 
The key idea is to leverage the \emph{intrinsic dissipation} responsible for gate errors in one ($U_3(\theta,\phi,\lambda)$) and two qubit (CNOT) gates. 
The precise details of this dissipation are not known and surely dependent on the specific machine and qubit(s). 
We can however assume \emph{a priori} that the dissipation is limited to one or two qubit processes (i.e. that it is \emph{local}) and otherwise \emph{generic}, an assumption we confirm a posteriori by the very good match to theory.

The Lindblad equation for the evolution of the density matrix $\rho(t)$ is then spelled out 
\begin{equation}
\partial_t \rho(t) = \sum_{n,m=1}^{N_\ell} K_{nm} \left( L_n \rho L_m^\dagger - \frac 1 2 \left\{ L_m^\dagger L_n, \rho \right\}\right).
\end{equation}
The set of operators $\{L_n\}$, $n=1\dots N_\ell$ describes all possible dissipation channels we consider. For the case of \emph{one body dissipation}, it is limited to operators acting on a single qubit 
\begin{equation}
\begin{split}
  L_1 &= I \times I \times \dots \times I \times X \\
  L_2 &= I \times I \times \dots \times I \times Y \\
  &\vdots\\
  L_{3\ell} &= Z \times I \times \dots \times I \times I, \\
  \end{split}
\label{eq:onebodylindblads}
\end{equation}
where $X$, $Y$ and $Z$ denote Pauli matrices and $I$ is the one qubit identity operator.

For the case of \emph{two body dissipation}, it includes in addition all two
qubit operators (which act on nearest neighbors in the qubit geometry given by
the CNOT connectivity of the quantum processor). Given an open 1-D
chain topology, which is what we will be using, there are a total of $N_\ell=3\ell
+ 9(\ell-1)$ operators, while for the completely connected case, there are $N_\ell=3\ell +9\ell(\ell-1)/2$
\begin{equation}
\begin{split}
L_{3\ell+1} &= I \times I \times \dots \times I \times X \times X \\
&\vdots\\
L_{N_\ell} &= Z \times Z \times I \times \dots \times I \times I. \\
\end{split}
\label{eq:twobodylindblads}
\end{equation}

The strength and interaction between the dissipation channels described by the operators $L_n$ is included in the Kossakowski matrix $K_{nm}$. We consider here completely generic one and two body dissipation, which means that we model $K_{nm}$ by a positive semidefinite random matrix. 

Surprisingly, for this case one can make detailed theoretical predictions \cite{wang_hierarchy_2020} for the relative decay rates of different observables, i.e. the \emph{many-body coherence times}. In this paper, we compare the theoretical prediction to the results of our quantum simulation experiment on the IBM machines.

In particular, the decay rates of $k$ qubit observables $\Tr\left[ \rho_0 O^{(k)}(t) \right]$ (i.e. observables $O^{(k)}$ which act on $k$ qubits, and are identity on the remaining $\ell-k$ qubits)  depend strongly on $k$ \cite{wang_hierarchy_2020}. Specifically for the case of one body dissipation, the decay timescales obey $1/\tau_k \propto k$. In the case of two body dissipation, one obtains a quadratic polynomial $1/\tau_k \propto -2k^2 +3 k\ell - k$, \emph{independently on the microscopic details} encoded in $K$ matrix, and for typical initial states $\rho_0$.
Our goal is to put this prediction to a rigorous experimental test using quantum simulation.

The general simulation strategy consists of the following ingredients:
\begin{itemize}
	\item Initialize qubits to initial product state $\ket{\psi_0}=\ket{\sigma_1^{(\gamma_1)}\sigma_2^{(\gamma_2)}\dots\sigma_\ell^{(\gamma_\ell)}}$, where each qubit can be in a different local basis $\gamma_i\in\{x,y,z\}$. If needed, we rotate to the corresponding basis using Hadamard and S gates.
	\item ``waiting'' circuit of depth $t$ to delay measurement till ``time'' $t$ (measured in gate times) and allow dissipation to act.
	\item Measurement of a $k$ body observable $O^{(k)}$ given as a Pauli string with $k$ nonidentity operators (again rotating to the corresponding local basis as necessary).
\end{itemize}
We use different designs of ``waiting'' circuits which introduce either one body (i.e. Lindblad operators from Eq. \eqref{eq:onebodylindblads}) or two body dissipation to the system (i.e. Lindblad operators of the form \eqref{eq:twobodylindblads}).

\subsubsection{One body dissipation}
\begin{figure}[t]
	\includegraphics[width=0.6\columnwidth]{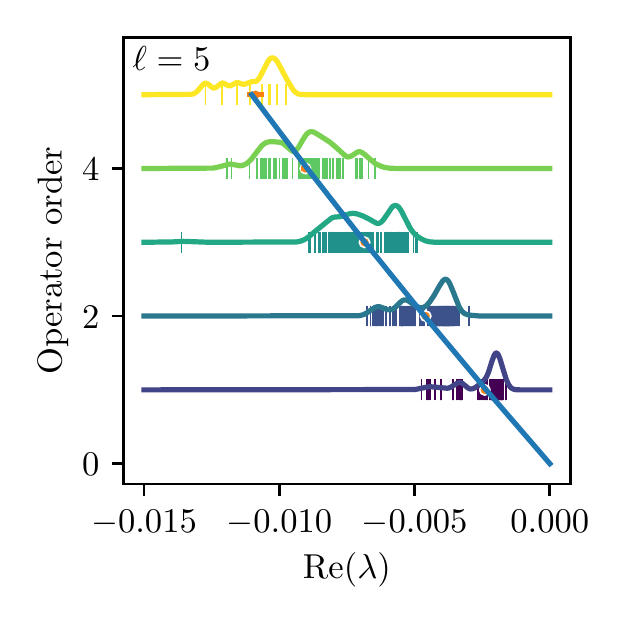}
	\caption{Real part of the eigenvalues $\lambda$ of the Liouvillian $\mathcal L$ extracted from 1 body dissipation circuits $W_1$ on \texttt{ibmq\_bogota}. The average inverse timescales for each order $k$ align with predicted linear hierachy, where the operator order $k$ refers to the number of non-identity pauli operators in the operator. The smooth curves show the density profile of timescales for each order.
		\label{fig:exp_1body}  }
\end{figure}
\label{sec:1body}
To study generic one body dissipation, we use a waiting circuit of depth $t$, consisting of random one qubit
gates, see Fig \ref{fig:1body_circuit}. The circuit $W_1^{(t)}$ consists of a subcircuit $C_1^{(t)}$ of $t$ layers of random single qubit unitaries $U_3(\theta,\phi,\lambda)$. 
It is followed by the inverse ${C^{(t)}}^\dagger$, undoing the unitary part of the action of $C_1^{(t)}$. Therefore, the total action of the waiting circuit $W^{(t)}$ would be identity for perfect gates. Due to the imperfections of the quantum computer, however, a non-unitary part remains due to dissipative processes. 

We note that the dissipation introduced by gate errors in this protocol is much stronger than residual two qubit interactions in the computer, which can therefore be neglected, hence yielding a purely dissipative system, which we model by a random 1-local Liouvillian. The theoretical prediction \cite{wang_hierarchy_2020} for this case is that the decay timescales $\tau_k$ of $k$ body observables $O^{(k)}$ should be proportional to $1/k$, which means that the real parts of the corresponding eigenvalues $\lambda$ of the Liouvillian should scale as $\text{Re}\left(\lambda^{(k)}\right) = 1/\tau_k \propto k$.

Fig. \ref{fig:exp_1body} displays these real parts of the eigenvalues,
reconstructed from measured decays of $k$-qubit Pauli string observables using
harmonic inversion (cf. Sec. \ref{sec:methods}), organized by the operator order
$k$ for a large number of observables in the $5$ qubit machine
\texttt{ibmq\_bogota}. A clear hierarchy is visible, revealing that more complex
many-qubit operators decay faster than less complex ones. While the distribution of reconstructed eigenvalues reveals some fine structure, the average inverse decay timescale (orange dots) show a clear linear scaling with $k$ (blue curve). This confirms the behavior predicted from theory.

\subsubsection{Two body dissipation}
\label{sec:2body}
\begin{figure}[t]
	\includegraphics{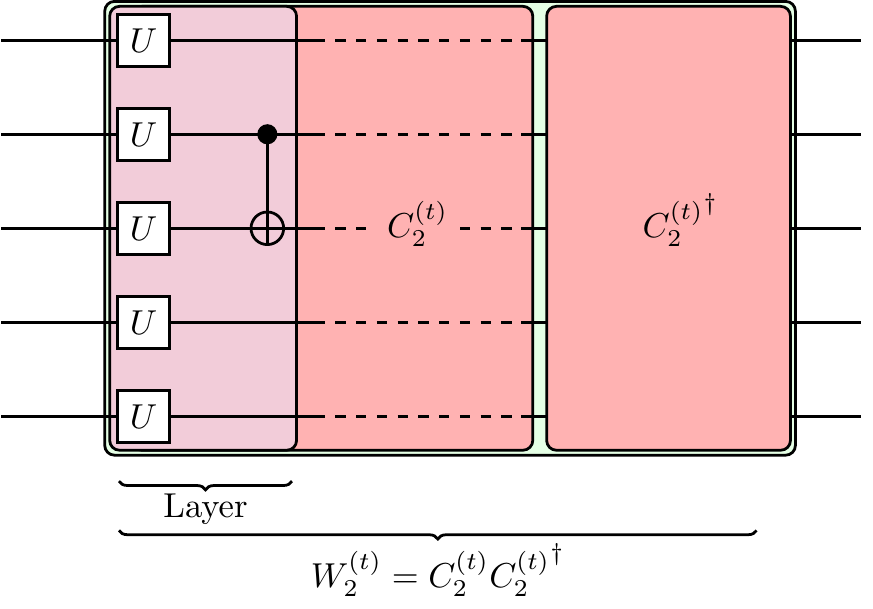}
	\caption{Example of waiting circuit $W_2^{(t)}$ introducing two qubit dissipation. It has the same
		structure as for one body dissipation see Fig. \ref{fig:1body_circuit}, except each layer now also contains one CNOT gate acting between a random pair of nearest neighbours according to the machine topology.\label{fig:2body_circuit} }
\end{figure}

\begin{figure*}[t]
	\includegraphics[width=\textwidth]{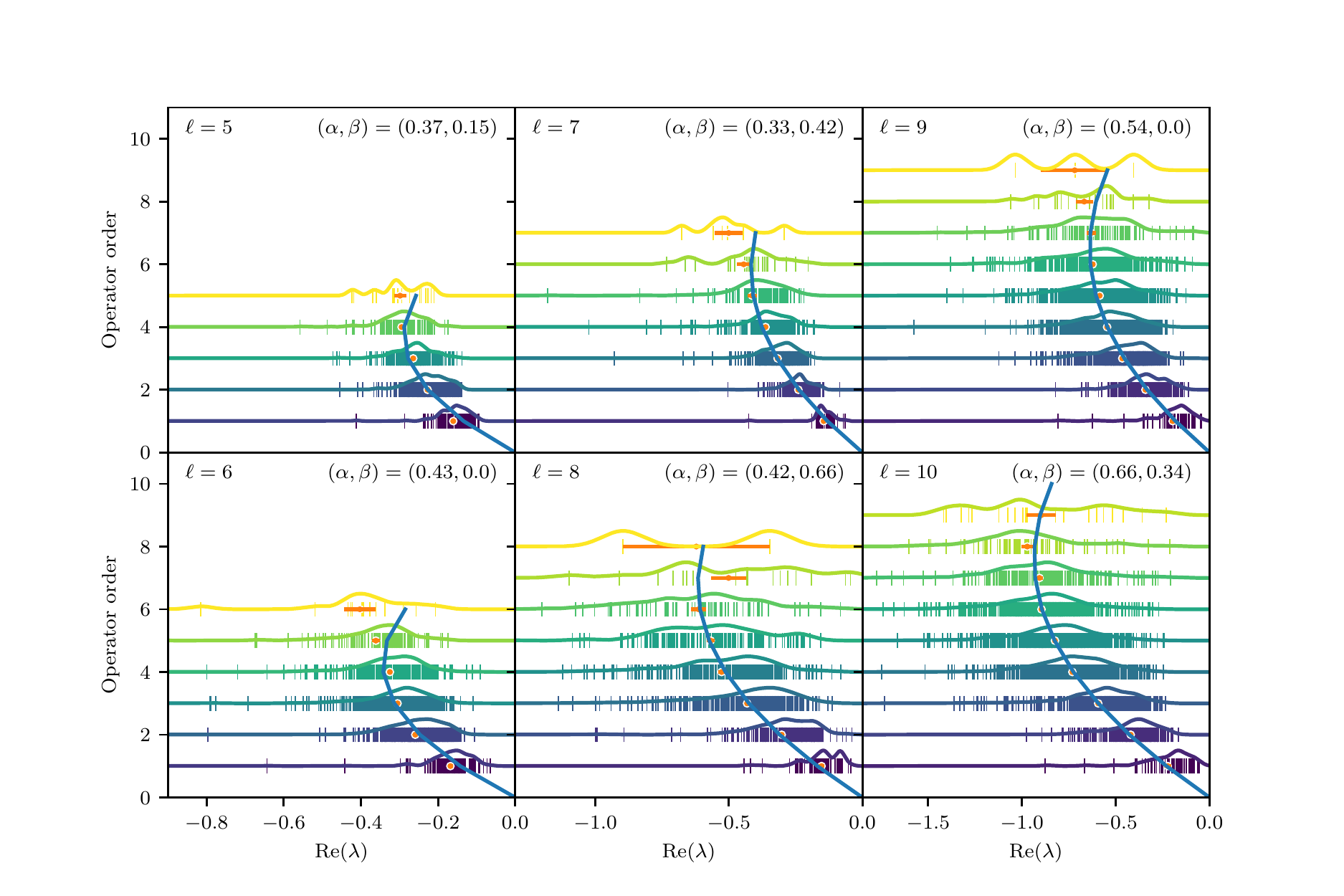}
	\caption{ Real part of eigenvalues $\lambda$ (vertical bars) of the 2-local Liouvillian 
		reconstructed from many-body operator decay time traces using the two body dissipative waiting circuit $W_2^{(t)}$. 
		The reconstructed eigenvalues are organized by their complexity or operator order (number of nonidentity operators $k$) of the observables and their density is indicated by the colored lines.
		The simulations were performed on \texttt{ibmq\_16\_melbourne}, using different subsets of $\ell$ qubits to explore a range of system sizes.
		For each $k$, the orange dots depict the average inverse decay timescale.
		The blue curve corresponds to a two parameter fit ($\alpha$, $\beta$) of the theoretical prediction from Eq. \eqref{eq:2bodyscales} for the decay rates. 
		\label{fig:turnback} }
\end{figure*}

\begin{figure*}[t]
	\includegraphics[width=\textwidth]{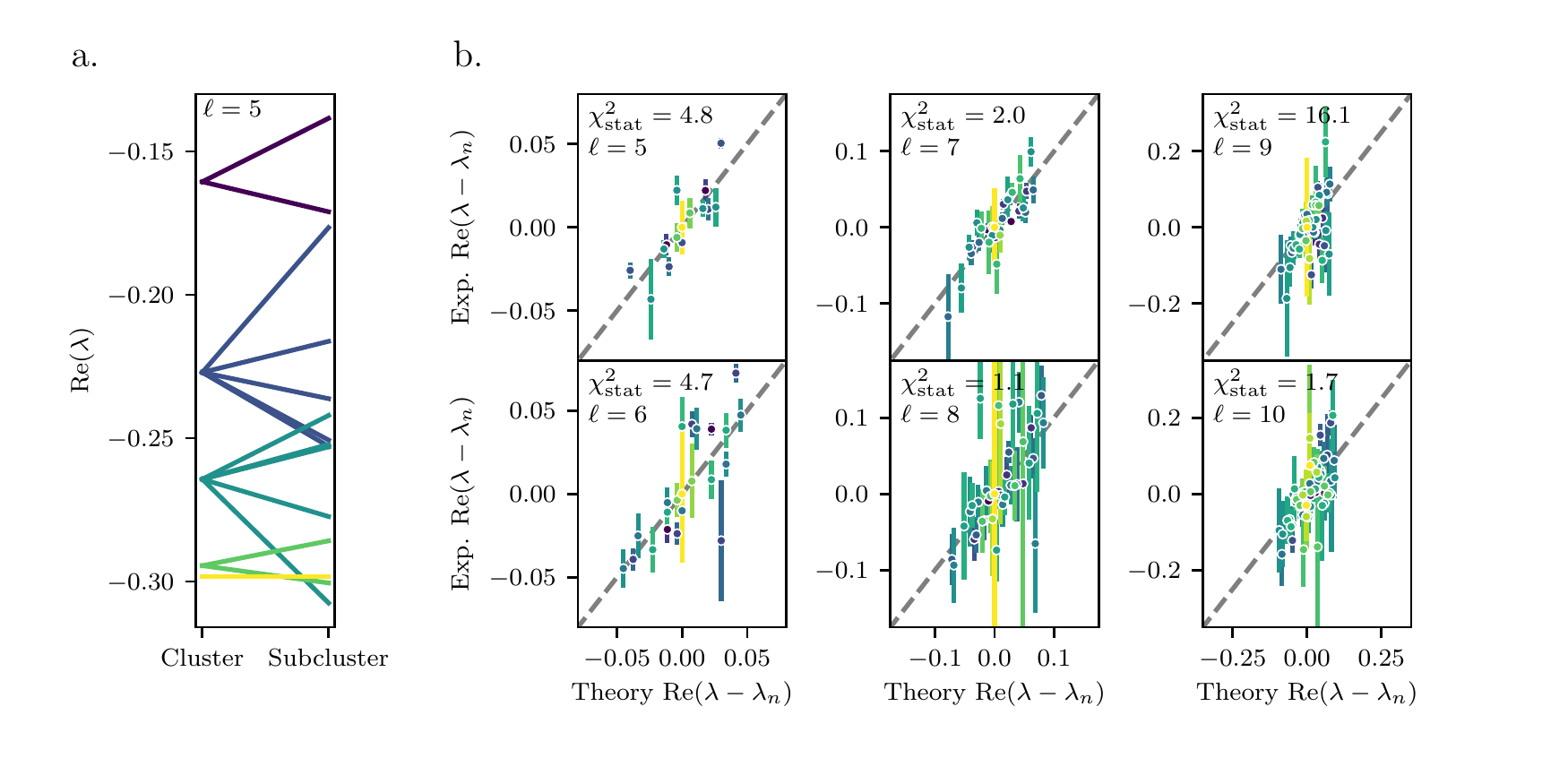}
	\caption{The eigenvalue clusters based on operator order are split due to the spatial
		structure of the two-body dissipation. These subclusters are indexed by operator
		order $k$, number of pairs of non-identities and number of nonidentities on edge sites. 
		Subfigure a shows the degree the $\ell=5$ clusters are split on
		\texttt{ibmq\_16\_melbourne}. The dissipation only acts on an chain subset of $\ell$ qubits with open boundary conditions, on which we include CNOT gates in our waiting circuit $W_2(t)$.
		Subfigure b is the distance between the subcluster centers and the cluster
		centers observed in \texttt{ibmq\_16\_melbourne} and in theory. We observe qualitative
		agreement for all, but the normalised $\chi^2$ based on the statistical error
		$\chi^2_\mathrm{stat}$ is very large for some experiments. The large
		$\chi^2$ is primarily caused by a few data points with small variance, and removing
		those brings $\chi^2_\mathrm{stat}$ below $2$ for all experiments. \label{fig:seperation}}
\end{figure*}

Next, we consider the more complicated case of generic two body dissipative interactions. 
In order to introduce predominantly two body dissipation, we leverage gate errors of the two qubit CNOT gate.
We design again a two component waiting circuit (cf. Fig \ref{fig:2body_circuit}) $W_2^{(t)}=C_2^{(t)} {C_2^{(t)}}^\dagger$, which contains random single site unitaries $U_3(\theta,\phi,\lambda)$ to introduce one qubit dissipation as well as to rotate to a random on-site basis, and CNOT gates between random pairs of neighbors allowed by the qubit geometry. Again, the application of the inverse circuit ${C_2^{(t)}}^\dagger$ removes all unitary components, and we are thus left with a purely dissipative system with dissipation processes corresponding to the gate errors of one qubit and two qubit gates.

This can be modelled with a random 2-local Liouvillian \cite{wang_hierarchy_2020}.
For random 2-local Liouvillians, the spectrum of the Liouvillian splits into distinct eigenvalue clusters, each of which governs the 
decay of $k$-qubit observables. The real parts of the eigenvalue clusters correspond to inverse decay timescales
of $k$-qubit observables (consisting of superpositions of Pauli strings with $k$
nonidentities and $\ell-k$ identities) and can be calculated theoretically \cite{wang_hierarchy_2020} to yield
\begin{equation}
    1/\tau_k= \frac{\alpha}{9(\ell-1)}(3\ell k - 2k^2 -k) + \frac{\beta}{3\ell} k
\label{eq:2bodyscales}
\end{equation}
where the two parameters $\alpha,\beta$ correspond to the relative strengths of two and one-body
dissipation processes respectively (the normalisation of $\alpha$ and $\beta$
count the number of relevant two and one body dissipation channels for an open 1-D
chain topology).

Surprisingly, there is a nonmonotonic behavior as a function of observable complexity $k$: The fastest decay rate is found 
for $k_*\approx (3(\ell-1)\beta+\ell(3\ell-1)\alpha)/{4\ell\alpha}$, and more complex observables with $k>k_*$ decay \emph{slower} than this maximal rate.
We call this feature of decay timescales \emph{turnback}, which is characteristic for dissipative interactions, and not present in simple 
one body dissipation.

As for the case of one body dissipation, we again perform simulations to measure the decay of a large number of Pauli string operators $O^{(k)}$ with $k$ nonidentity Pauli matrices under the action of the waiting circuit $W_2^{(t)}$. From these time traces, we reconstruct the real parts of the Liouvillian eigenvalues using harmonic inversion (cf. \ref{sec:methods}) and compare them to the theoretical prediction in Fig. \ref{fig:turnback}. Since the overall strength of the dissipation is not known a priori, we fix the parameters $\alpha$ and $\beta$ by a fit, yielding excellent agreement with the positions of the centers of the eigenvalue clusters obtained from experiment. Most strikingly, the \emph{turnback} of decay rates with increasing operator order is well reproduced in the quantum simulation. We note that different subsets of size $\ell$ of \texttt{ibmq\_16\_melbourne} include different qubits and the CNOT and one qubit gate errors depend on the specific choice of qubits, which explains the variance in the fit parameters $(\alpha,\beta)$ for different $\ell$.

The reconstructed eigenvalues show some fine structure within a given operator order, which is revealed by the eigenvalue density curves.
It is important to note that all IBM quantum computers have a notion of qubit distance, which limits the possible combinations of qubits by CNOT gates.
Therefore, our simulation actually corresponds to a \emph{spatially local} two qubit dissipative interaction, where the spatial connectivity is given by the qubit architecture. Including this structure in the theoretical description leads to further splitting of each eigenvalue cluster of order $k$ into subclusters at different positions (cf. Fig. \ref{fig:seperation}), based on the number of neighboring nonidentity pairs $p$ in the operator string and the number of nonidentities on edge sites $e$. As explained in Sec. \ref{sec:PT}, we therefore get to leading order in perturbation theory eigenvalue clusters labeled by $(k,p,e)$, depending on the connectivity of the qubit subset used in the simulation.
These corrections destroy the perfect separation of eigenvalue clusters and are the reason why different order $k$ decay rates overlap in our experimental results in Fig. \ref{fig:turnback}. The overall hierarchy of the average decay rates in terms of the number of nonidentities $k$ in the observables however remains.

Although the experimental precision of eigenvalue reconstruction is limited, we compare the predicted corrections to the decay rates for a subset of $\ell$ qubits of \texttt{ibmq\_16\_melbourne} arranged as a linear chain with open boundary conditions in Fig. \ref{fig:seperation}b. This is achieved by including only CNOT and $U_3$ gates on the chain in our waiting circuit $W_2(t)$.
For each predicted ``subcluster'' of eigenvalues of $\mathcal L$, we calculate its deviation from the original location both for our experimental and theoretical results and plot them against each other. Despite relatively large errorbars from the experiment, these results appear to be statistically significant and confirm that the fine structure in decay rates can indeed be explained by the qubit connectivity of the machine, an observation which holds for different choices of machine subsets as shown in Fig. \ref{fig:seperation}b.

\section{Conclusion} 

We have demonstrated a simple and powerful way to simulate generic dissipative systems on current noisy quantum computers. The protocol leverages gate errors of one-qubit and two-qubit gates to generate single qubit dissipation as well as dissipative two qubit interactions. Our circuits are designed such that the unitary part completely cancels out, so that during the action of the circuit only an effective many-body dissipator remains. This protocol can be easily extended to simulate Hamiltonian systems in the presence of generic dissipation of this form, by reducing the amount of cancellation of the unitary part.
While the microscopic form of the introduced dissipation can not be controlled, generic features, such as the locality of dissipation channels can be selected at will. 
The simulation therefore corresponds to a fully generic, local, dissipative quantum many-body system, which we recently described using a random matrix theory \cite{wang_hierarchy_2020}. Our experimental results on the IBM quantum computer provide evidence for the emergence of a hierarchy of dissipative timescales, based on the $k$-body nature of observables, with a nontrivial, nonmonotonic behavior of the timescales as a function of $k$ in the case of dissipative interactions. This observation is in excellent agreement with the theoretical prediction, including fine structure caused by the qubit connectivity in the quantum processor.

These results put forward current quantum computers as ideal platforms
for studying the physics of generic quantum many-body systems, thereby
opening a new avenue in the field of digital quantum
simulation. This should be particularly relevant for the growing
community investigating dissipative quantum chaos.

\section{Methods}
\label{sec:methods}

\subsection{Experiment}

\subsubsection{Simulation protocol}

  Here we provide details on our implementation of generic open quantum systems in the IBM quantum computer. 
    
  \begin{figure}[pth]
  	\includegraphics[width=0.9\columnwidth]{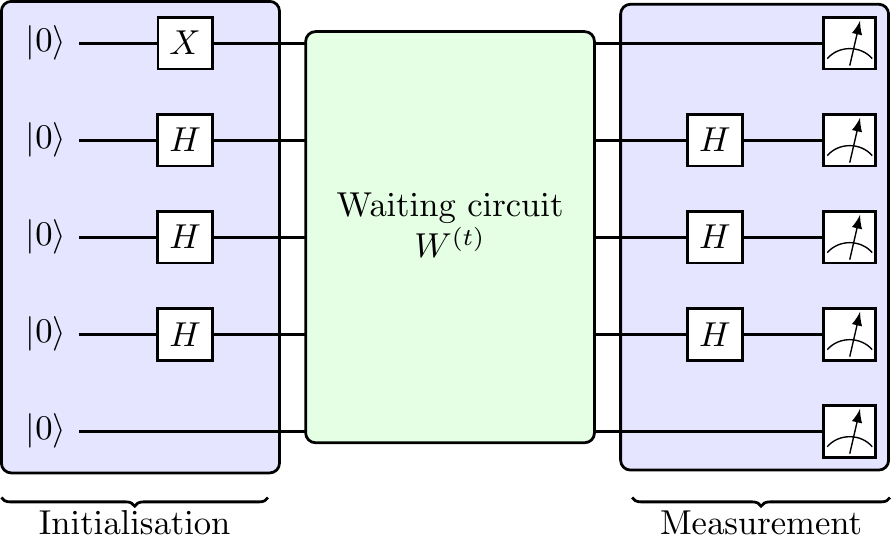}
  	\caption{The circuit we use to measure the decay rate of operators in a purely dissipative quantum system is made of three components. (i) The initialisation stage, where a product
  		state is created, rotating locally to the appropriate basis ($x$, $y$, or $z$). 
  		(ii) A waiting circuit of a variable depth $n$ (corresponding to simulated time), 
  		which is where the dissipation acts on the system. (iii)
  		A measurement stage, with appropriate rotations to the computational basis followed by
  		a measurement of each qubit in this basis, allowing us to find the evolution of
  		expectation values of up to $2^\ell-1$ different operators as we vary $t$.
  		\label{fig:general_circuit}}
  \end{figure}

  The general form of circuits we use is shown in Fig. \ref{fig:general_circuit}. Our goal is to measure 
  $\Tr\left(\rho_0 O^{(k)}(t)\right)$ for a large array of $k$-body observables of different degree of locality $k$ for different initial states $\rho_0$. For simplicity, we use simple pure product states $\ket{\psi_0} = \ket{\sigma_1^{\gamma_1}\dots\sigma_\ell^{\gamma_\ell}}$ (yielding $\rho_0 = \ket{\psi_0}\bra{\psi_0}$, where $\gamma_i\in \{x,y,z\}$ denotes the local basis and $\sigma_i\in [0,1]$ are the quantum numbers of the corresponding Pauli operator $ \sigma_{\gamma_i}$.
  
  After preparation of the initial state $\ket{\psi_0}$, we run different ``waiting circuits'' of depth $t$, which have the purpose of letting a generic local purely dissipative Liouvillian $\mathcal L$ act on the qubits for a time $t$. For nontrivial waiting circuits, we work in units of time given by the depth of the circuits, i.e. we will have integer time in units of the average gate time.
  For small $k$, (i.e. few body Pauli strings), we measure all strings made of $I$, $X$, $Z$, i.e. for each depth $t$ of the waiting circuit, we perform 8192 shots to measure $\Tr\left({\rho_0 O^{(k)(t)}}\right)$. 
  
  We then analyze these time traces using \emph{harmonic inversion} to decompose them into their contributions from complex exponential functions of the form 
  \begin{equation}
  \Tr\left(\rho_0 O^{(k)(t)} \right) = \sum_n \E^{\lambda_n t} c_n,
  \end{equation}
  where $\lambda_n\in\mathbb{C}$ correspond to the \emph{eigenvalues} of the effective adjoint Liouvillian governing the time evolution of the operators.
  
  \subsubsection{Harmonic inversion to extract Liouvillian eigenvalues from time traces}
  
    To decompose the experimental time traces $ \Tr\left(\rho_0 O^{(k)(t)} \right)$ into sums of complex exponentials $\E^{\lambda_n t}$, we use the harmonic inversion algorithm by Mandelshtam and Taylor \cite{mandelshtam_harmonic_1997}, which extends traditional filter diagonalisation methods. Its distinct advantage
    lies in the possibility of extracting both oscillation frequencies
    $\Im{\lambda_n}$ and decay rates $\Re{\lambda_n}$, without the frequency
    precision being limited by the total observation period of the signal as it
    would be with Fourier transformation. In the present
    case, this is useful as the decaying nature of our signals prevent long time
    observation.

    By choosing a small frequency interval of interest, and a maximal number of
    eigenvalues to look for in this interval, we can turn the problem of finding
    $c_n,\lambda_n$ into small matrix singular value decomposition problem. The maximal
    number of eigenvalues should be much greater than the expected number, and is
    limited only by the information content of the signal; from $t$ time steps,
    one can at most extract $\sim t/4$ complex pairs $c_n,\lambda_n$. Noise
    will shift the true eigenvalues slightly and create spurious modes. 
    The spurious modes can largely be identified based on a combination of their low amplitudes,
    distinct position, and an error-metric for $\lambda_n$ arising in the SVD problem. 

    The operators considered for the one-body and two-body dissipation
    experiments are dominated by eigenvalues closer to a point on the
    real axes. The reason for this is the following: The natural dissipation channels (given by eigenmodes of the Kossakowski matrix $K$) can be expected to be essentially expressed in a random (local) basis, while we measure the decay of Pauli string observables in the $x,y,z$ basis. Due to this basis missmatch, we find that each observable sees contributions of essentially all eigenvalues of $\mathcal{L}$ corresponding to a certain locality class $k$, which have different imaginary parts, but very similar real parts due to the timescale hierarchy. This means that oscilating contributions essentially cancel due to interference, leaving only decaying contributions, with decay timescales ranging over the width of the corresponding eigenvalue cluster.    
    
    It is therefore theoretically possible to use exponential
    regression to extract the decay rate of a time trace. We choose to use
    harmonic inversion as it generalises to the case where the imaginary part of $\lambda_n$ is significant, and in practice we find better agreement between theory and experiment than with regression. We believe this is because  harmonic inversion is better able to filter out spurious modes in 
    noisy time traces.

\begin{figure}[pth]
    \includegraphics[width=\columnwidth]{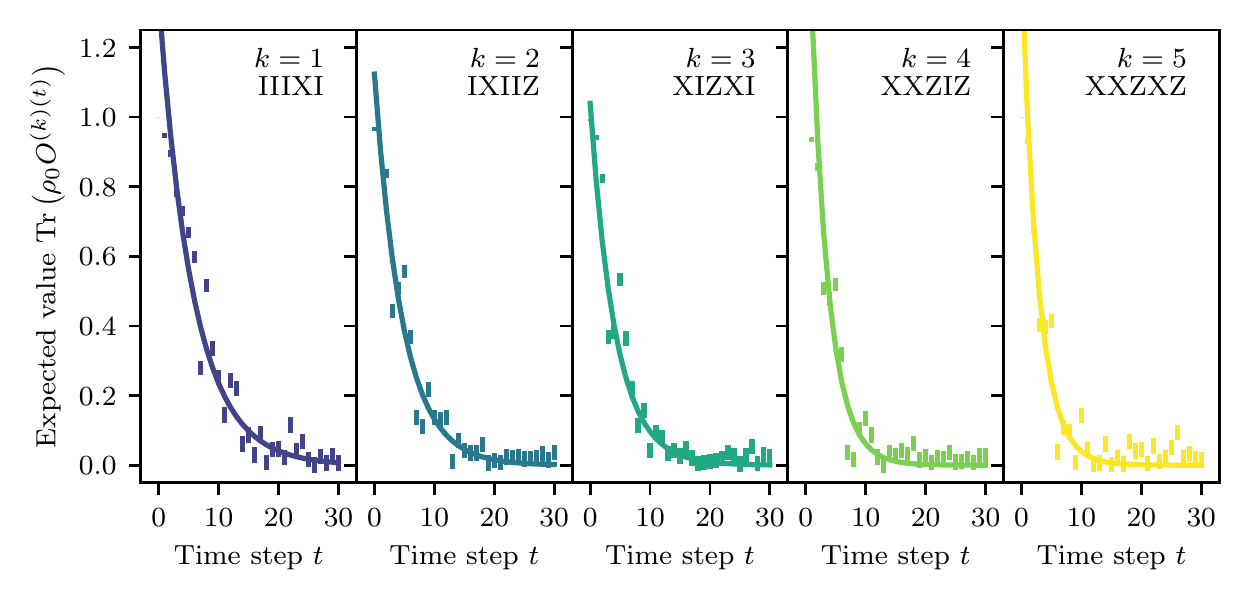}
    \caption{Generic time traces of $k$-body operators in the $\ell=5$
        \texttt{ibmq\_16\_melbourne} (vertical bars) superposed with the predicted time traces from their
        harmonic inversion decompositions. Good agreement is seen,
        which justifies our use of harmonic inversion.  \label{fig:timetrace}}
    \end{figure}

\begin{figure}[pth]
    \includegraphics[width=0.9\columnwidth]{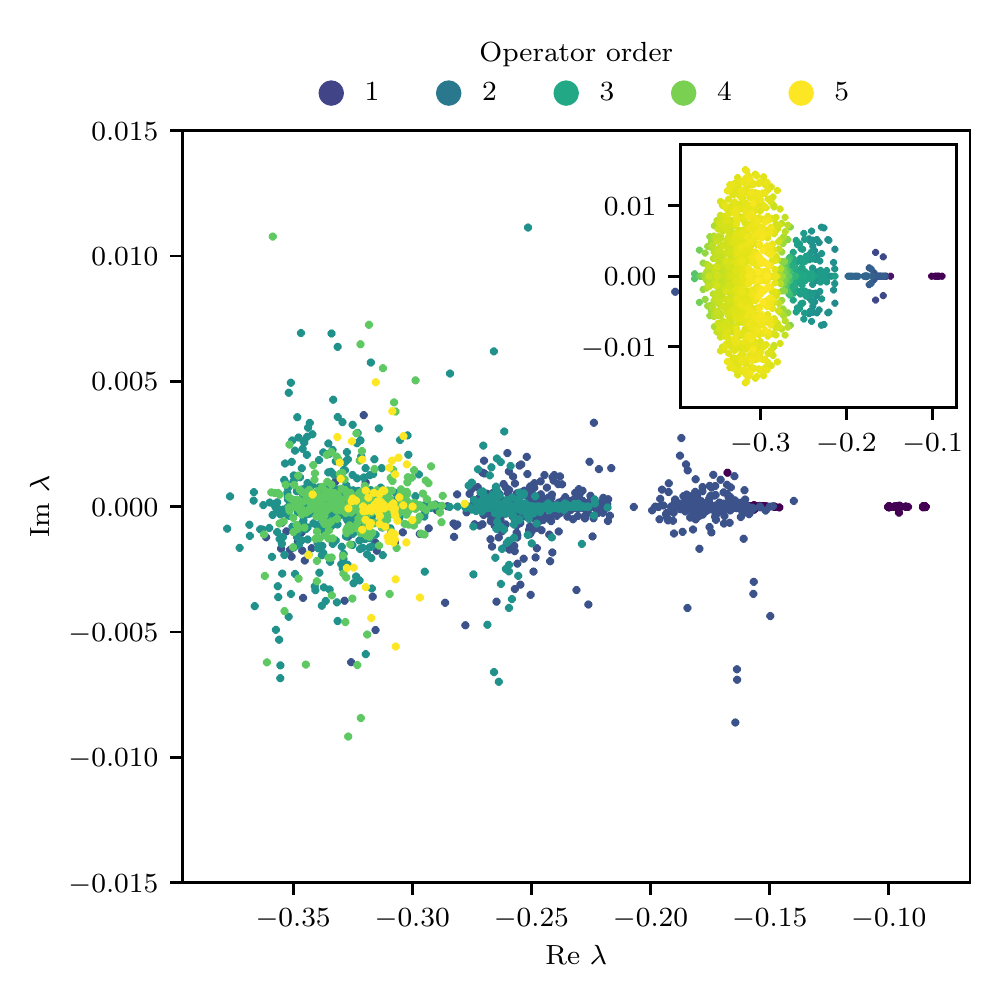}
    \caption{Eigenvalues extracted by harmonic inversion from ~$1000$ simulated time traces $\tr(\rho_0 {O}^{(k)})$ of a two-body spatially local $\ell=5$ Liouvillian
        $\mathcal{L}$. The inset shows the true spectrum, obtained by exact
        diagonalisation. It is colored by the average operator order of the
        eigenoperators of the adjoint Liouvillian $\mathcal L^\dagger$. The initial states $\rho_0$ and
        operators ${O}^{(k)}$ are chosen in a similar fashion to our experimental procedure. The eigenvalues are
        colored by the order $k$ of the operator of the decomposed time
        trace. We see good agreement in terms of the overall
        $\mathrm{Re}\,\lambda$ structure,
        and closer investigation reveals that various cluster centers due to
        locality and operator order also agree well. The slight 
        between the two spectra arise due to some eigenvectors of $\mathcal L$ having significant contributions from multiple operator orders. The imaginary structure is reproduced
        less clearly, since it is orders of magnitude smaller than the real part, and so has only a small impact on time dynamics, and also due to the basis missmatch between the eigenoperators of $\mathcal L^\dagger$ and Pauli string observables as explained in the text.  
        We can still extract the scale of the imaginary part of the spectrum. 
    \label{fig:reconstruct}}
    \end{figure}

    As harmonic inversion is not standard in the literature, we supply evidence
    that it is suitable in Figs. \ref{fig:timetrace} and
    \ref{fig:reconstruct}. In particular, they show that harmonic inversion can reproduce
    the spectrum of a two-body spatially local Liouvillian from simulated time
    traces, and that there is good agreement between the experimentally observed
    time traces, and the time traces their harmonic inversion decomposition predicts.

\subsection{Theory}

\subsubsection{Perturbation theory}
\label{sec:PT} We use nonhermitian degenerate perturbation theory detailed in the main text and
supplementary material of Ref. \cite{wang_hierarchy_2020}. In a nutshell, we
consider a purely dissipative $\ell$ qubit system, governed by a set of Lindblad
operators $\{L_i\}$, $i=1,\dots,N_\ell$. The adjoint Liouvillian $\mathcal
L^\dagger$ is given by 
\begin{equation}
	\mathcal{L}^\dagger[\bullet] = \sum_{n,m=1}^{N_\ell} K_{n,m} \left( L_m^\dagger \bullet L_n - \frac 1 2 \left\{ L_m^\dagger L_n, \bullet \right\} \right).
\end{equation}
Since $K$ needs to be a positive semidefinite matrix, we sample it using a diagonal random matrix $D$ with nonnegative eigenvalues $d_i$ and a random CUE matrix $U$ sampled from the Haar measure, to yield $K=U^\dagger D U$. We normalize $\tr{K}=N=2^\ell$, and it turns out that $K$ is diagonal dominant with
on average $\text{mean}(K_{nn})=N/N_\ell=d$. Therefore, we can approximate $K=dI+ K'$, with a small perturbation $K'$ \cite{wang_hierarchy_2020}.
If we approximate $K$ by $dI$ ($I$ is the identity), we obtain the starting point for perturbation theory, and can analyze the block structure of the adjoint Liouvillian. We use for convenience the basis of normalized Pauli string operators
\begin{equation}
	S_{\vec{x}} = \frac{1}{\sqrt{N}} \sigma_{x_1} \times \sigma_{x_2} \times \dots \sigma_{x_\ell},
\end{equation}
where $x_i\in\{0,1,2,3\}$, and can calculate the diagonal elements (offdiagonals vanish) of the adjoint Liouvillian in this basis:
\begin{equation}
	\mathsf{L}_{xx} = \tr\left( S_{\vec{x}} \mathcal{L}^\dagger[S_{\vec{x}}] \right)=
	\sum_{n}^{N_\ell} d \tr \left( S_{\vec{x}} L_n S_{\vec{x}} L_n - \frac 1 N I \right).
\end{equation}
Here, we note that $L_n$ are Pauli strings and therefore if $S_{\vec{x}}$ and $L_n$ commute, we get zero, since Pauli matrices square to identity. It is now easy to see that we get ``blocks'' of identical $L_{xx}$ under certain conditions. (i) If the $L_n$ are \emph{all} only one body operators, we get the same $L_{xx}$ for all strings $S_{\vec{x}}$, which have the same number of nonidentities.
(ii) If $L_n$ furthermore include \emph{all} two body operators, we get the same condition, but different diagonal matrix elements of the adjoint Liouvillian.
(iii) If there is a spatial structure in the two body Lindblad operators, allowing only nearest neighbor operators, we get different ``blocks'' based on the number of nonidentities $k$ in $S_{\vec{x}}$, the number of edge nonidentities (since edge sites have less neighbors), and the number of neighboring nonidentities.

For $K=dI$, we therefore get a diagonal matrix representation of $\mathcal L^\dagger$, which is the starting point of our perturbation theory, yielding degenerate eigenvalues given by the diagonal elements $\mathsf{L}_{xx}$. Now, we include the offdiagonal part of $\mathcal L^\dagger$, generated by $K'=K-dI$ as a perturbation, which lifts the degeneracy of the eigenvalue clusters. 

\subsubsection{Exact diagonalization of $\mathcal{L}$}

   We use exact diagonalization of model Liouvillians on few qubit systems. For this, we build the adjoint Liouvillian superoperator as a matrix in the many-body operator Hilbert space of dimension $4^\ell$ for $\ell$ qubits, spanned by all possible Pauli strings. We then diagonalize this matrix using standard \texttt{lapack} routines to find its eigenvalues and eigenvectors.

\begin{acknowledgments}
  We thank IBM for access to their quantum computers through  IBM Quantum Experience for Researchers. This work was in part supported by the Deutsche Forschungsgemeinschaft (DFG) through SFB 1143 (project-id 24731007).
\end{acknowledgments}

\bibliography{qchierarchy.bib}
  
\end{document}